\documentstyle[11pt,newpasp,twoside,psfig,graphicx]{article}
\def\edcomment#1{\iffalse\marginpar{\raggedright\sl#1\/}\else\relax\fi}
\reversemarginpar

\newcommand{\ltsima}{$\; \buildrel < \over \sim \;$}
\newcommand{\simlt}{\lower.5ex\hbox{\ltsima}}
\newcommand{\gtsima}{$\; \buildrel > \over \sim \;$}
\newcommand{\simgt}{\lower.5ex\hbox{\gtsima}}

\def\lesssim{\mathrel{\hbox{\rlap{\hbox{\lower4pt\hbox{$\sim$}}}\hbox{$<$}}}}
\def\gtrsim{\mathrel{\hbox{\rlap{\hbox{\lower4pt\hbox{$\sim$}}}\hbox{$>$}}}}

\def\ab1450{$AB_{1450(1+z)}$}

\def\xray{\hbox{X-ray}}

\def\today{\ifcase\month\or January\or February\or March\or April\or May\or
      June\or July\or August\or September\or October\or November\or December\fi
      \space\number\day, \number\year}

\def\chandra{{\it Chandra\/}}
\def\conx{{\it Constellation-X\/}}
\def\genx{{\it Generation-X\/}}

\def\heao1{{\it HEAO-1\/}}

\def\rosat{{\it ROSAT\/}}

\def\xeus{{\it XEUS\/}}
\def\xmm{{XMM-{\it Newton\/}}}

\begin{document}
\title{X-rays from the Dawn of the Modern Universe. 
Chandra and XMM-Newton Observations of \boldmath$z>4$ Quasars}
\author{C. Vignali}
\affil{INAF - Osservatorio Astronomico di Bologna, Via Ranzani 1, 40127 
Bologna, Italy}
\author{W.N. Brandt, D.P. Schneider}
\affil{Department of Astronomy \& Astrophysics, The Pennsylvania State 
University, 525 Davey Lab, University Park, PA 16802, USA}

\begin{abstract}
Quasars at $z>4$ provide direct information on the first massive structures 
to form in the Universe. Recent ground-based optical surveys (e.g., 
the Sloan Digital Sky Survey) have discovered large numbers of high-redshift quasars, 
increasing the number of known quasars at $z>4$ to $\approx$~500. 
Most of these quasars are suitable for follow-up \xray\ studies. 
Here we review \xray\ studies of the highest redshift quasars, 
focusing on recent advances enabled largely by the capabilities of \chandra\ and \xmm. 
Overall, analyses indicate that the \xray\ emission and 
broad-band properties of high-redshift and local quasars are 
reasonably similar, once luminosity effects are taken into account. 
Thus, despite the strong changes in large-scale environment and 
quasar number density that have occurred from $z\approx$~0--6,  
individual quasar \xray\ emission regions appear to evolve relatively little. 
\end{abstract}

\section{Introduction}

Our knowledge of the \xray\ properties of quasars at $z>4$ has advanced rapidly 
over the past few years. 
In particular, the Sloan Digital Sky Survey (SDSS; e.g., York et al. 2000) has generated 
large and well-defined samples of $z>4$ quasars (e.g., Anderson et al. 2001); 
most of these quasars are suitable for \xray\ studies. 
The \xray\ observational strategy has comprised archival studies of 
high-redshift quasars with \rosat\ (Kaspi, Brandt, \& Schneider 2000; Vignali et al. 2001, 
hereafter V01), snapshot \hbox{($\approx$~4--10~ks)} observations with \chandra\ to define 
basic quasar \xray\ properties such as fluxes and luminosities 
(e.g., V01; Brandt et al. 2002, 2003; Bechtold et al. 2003; 
Vignali et al. 2003a,b, hereafter V03a, V03b) 
and longer observations with \xmm\ to derive 
either tight constraints on the \xray\ emission (e.g., Brandt et al. 2001) or 
spectral parameters by direct \xray\ fitting (Ferrero \& Brinkmann 2003; Grupe et al. 
2004). 
\chandra\ snapshot observations have also allowed joint spectral fitting of subsamples 
of quasars drawn from two main samples at $z>4$: 
the optically luminous Palomar Digital Sky Survey (e.g., Djorgovski et al. 1998) 
and the SDSS. The \xray\ spectral results provide 
no evidence of strong spectral evolution in radio-quiet quasar (RQQ) \xray\ emission 
from local samples up to \hbox{$z\approx$~5}; the spectrum at high redshift is 
well parameterized by a power law in the 
\hbox{$\approx$~2--40~keV} rest-frame band with \hbox{$\Gamma=1.8$--2} (V03a; V03b). 
Furthermore, no evidence for widespread intrinsic \xray\ absorption has been found, although it seems 
likely that a few individual objects may be \xray\ absorbed (e.g., V01; 
V03b). These overall results have been supported recently by direct 
\xray\ spectroscopy of QSO~0000$-$263 at $z=4.10$ with \xmm\ (Ferrero \& Brinkmann 2003). 

The color selection of the SDSS has been proven to be effective in finding high-redshift 
optically luminous quasars up to \hbox{$z\approx$~5.7} (see Fan et al. 2003 for SDSS quasars 
at higher redshifts). 
On the other hand, moderately deep \chandra\ observations and the ultra-deep (2\ Ms) survey of 
the \chandra\ Deep Field-North (CDF-N; Alexander et al. 2003) 
can detect Active Galactic Nuclei (AGN) at $z>4$ 
that are typically \hbox{$\simgt$~10--30} times less luminous than the SDSS quasars 
(e.g., Barger et al. 2002; Silverman et al. 2002; Vignali et al. 2002, 
hereafter V02; Castander et al. 2003). 
These AGN are much more numerous and therefore more representative of the AGN population at 
high redshift than the rare SDSS quasars; however, their \xray\ emission does not appear to 
contribute significantly to reionization at \hbox{$z\approx$~6} (Barger et al. 2003). 
A detailed \xray\ spectral analysis of the $z>4$ AGN in the CDF-N is presented in V02. 

Below we present some new \xray\ spectral results obtained by joint spectral fitting of 
all the RQQs at $z>4$ thus far detected by \chandra. 
A spectral analysis performed on a smaller but more \xray\ 
luminous sample of $z>4$ radio-loud quasars is presented 
by Bassett et al. (in preparation).

\section{Joint X-ray Spectral Results}
To define the overall \xray\ properties of $z>4$ RQQs, 
we selected all of the RQQs detected by 
\chandra\ with $>2$ counts in the observed \hbox{0.5--8~keV} band. The sample comprises 
46 quasars with a median redshift of 4.43; 
the number of source counts is \hbox{$\approx$~750}. 
Note that these quasars 
represent a large fraction 
($\approx$~70\%) of the optically selected RQQs at $z>4$ with \xray\ detections at present.\footnote{See 
http://www.astro.psu.edu/users/niel/papers/highz-xray-detected.dat 
for a regularly updated listing of \xray\ detections and sensitive upper limits at $z>4$.} 
Although it is possible that individual objects are characterized by ``peculiar'' 
\xray\ properties, our approach obtains average spectral 
parameters for the quasar population at $z>4$ 
using a much larger 
sample than those presented in V03a and V03b. 
In the \xray\ spectral analysis, 
the Cash statistic (Cash 1979) has been 
adopted. Our preliminary analysis shows that a power law fits the \xray\ data reasonably well; 
the photon index in the rest-frame \hbox{$\approx$~2--40~keV} band 
is $\Gamma=1.9\pm{0.1}$ (see Fig.~1). 
%
\begin{figure}[!h]
\centering\includegraphics[width=13cm,height=8.1cm,angle=0]{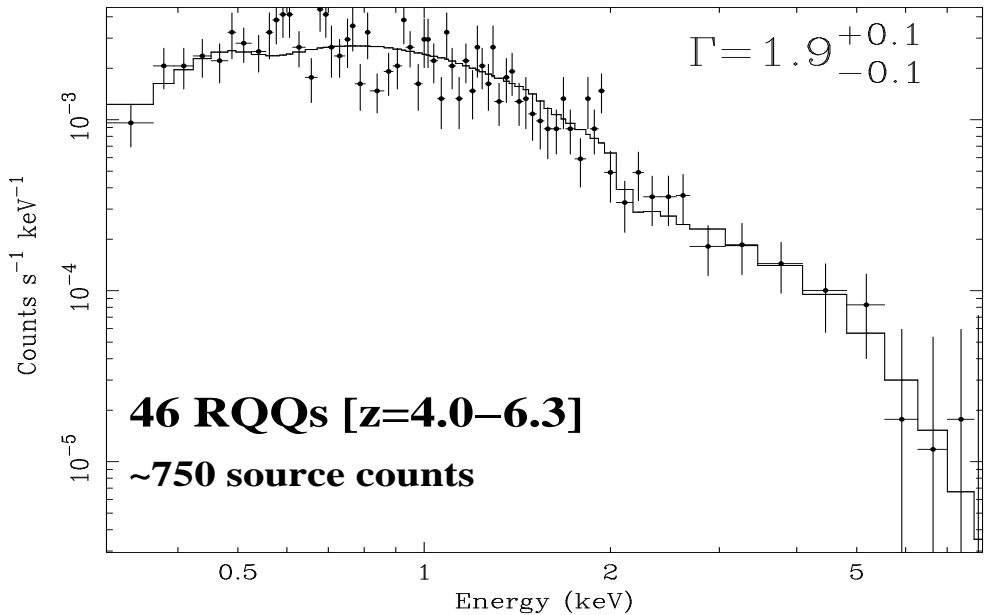}
\caption{Combined spectrum of $z>4$ RQQs detected by \chandra\ with 
$>2$ counts in the observed \hbox{0.5--8~keV} band. 
The spectrum shown here (only for presentation purposes) 
is fitted with a power-law model and Galactic absorption (see the text for details).}
\label{fig:fig1}
\end{figure}
%
This is consistent with previous results obtained for RQQ samples at high redshift 
observed with \chandra\ (V03a; V03b) and 
\xmm\ (Ferrero \& Brinkmann 2003; Grupe et al. 2004), as well as with 
quasar \xray\ spectral results 
at low and intermediate redshift (e.g., George et al. 2000; Page et al 2003). 
Our analyses indicate that the \xray\ spectral properties of $z>4$ RQQs and local RQQs are 
similar; the only significant differences have been found in their broad-band properties using the 
SDSS Early Data Release quasar catalog (Schneider et al. 2002) and are likely due to luminosity effects 
(Vignali, Brandt, \& Schneider 2003; see also Brandt, Schneider, \& Vignali, these proceedings). 
Thus, despite the strong changes in large-scale environment and quasar number density 
that have occurred from $z\approx$~0--6, 
individual quasar \xray\ emission regions appear to evolve relatively little. 
From the joint \xray\ spectral fitting we also find no significant evidence for 
absorption above the Galactic value; the upper limit in the source rest frame is  
\hbox{$N_{\rm H}\simlt9\times10^{20}$~cm$^{-2}$} (see Vignali et al., in preparation, for 
detailed discussion).

\section{The Future}
The correlation found between quasar AB magnitude at a rest-frame wavelength of 1450~\AA\ and 
the observed \hbox{0.5--2~keV} flux (e.g., V03b) is a powerful tool to select 
samples of $z>4$ quasars suitable for follow-up \xray\ observations. 
The combination of snapshot observations with \chandra\ and longer exposures with 
\xmm\ should continue to be highly effective in allowing 
the study of the overall \xray\ properties 
of quasars at high redshift. In the coming years, as the SDSS is completed and several 
thousand \chandra\ and \xmm\ archival observations become available to the scientific community, 
our knowledge of the broad-band properties of quasars at the highest redshifts will significantly increase. 
However, detailed \xray\ spectroscopic analyses of large samples of $z>4$ quasars must await 
the more distant future and \xray\ missions such as \conx, \xeus, and \genx.

\acknowledgments
Support from the Italian Space Agency under contract ASI I/R/073/01 (CV), 
NASA LTSA grant NAG5-13035 (WNB, DPS), NSF CAREER award AST-9983783 (WNB), 
and NSF grant AST-9900703 (DPS) is gratefully acknowledged.

\end{document}